# Thin Current Sheet Behind the Dipolarization Front

R. Nakamura[1], W. Baumjohann[1], T. K. M. Nakamura[2,1], E. V. Panov[2,1], D. Schmid[1], A. Varsani[1], S. Apatenkov[3], V. A. Sergeev[3], J. Birn[4], T. Nagai[5], C. Gabrielse[6], M. André[7], J. L. Burch[8], C. Carr[9], I. S Dandouras[10], C. P. Escoubet[11], A, N. Fazakerley[12], B. L. Giles[13], O. Le Contel[14], C. T. Russell[15], and R. B. Torbert[8,16]

[1]Space Research Institute, Austrian Academy of Sciences, Graz, Austria

[2]Institute of Physics, University of Graz, Graz, Austria

[3]St Petersburg State University, St Petersburg, Russia

[4]Space Science Institute, Boulder, CO, United States

[5]ISAS/JAXA, Sagamihara, Japan

[6]The Aerospace Corporation, Los Angeles, CA, United States

[7]Swedish Institute of Space Physics, Uppsala University, Uppsala, Sweden

[8]Southwest Research Institute, San Antonio, TX, United States

[9]Imperial College London, London, United Kingdom

[10]IRAP, Université de Toulouse / CNRS / UPS / CNES, Toulouse, France

[11]ESTEC/ESA, Noordwijk, Netherlands

[12]Mullard Space Science Lab., Dorking, United Kingdom

[13]NASA Goddard Space Flight Center, Greenbelt, MD, United States

[14]LPP, CNRS, Observatoire de Paris, Paris, France

[15]University of California Los Angeles, Department of Earth Planetary and Space Sciences, Los Angeles, CA, United States

[16]Univ New Hampshire, Durham, NH, United States

Corresponding author: Rumi Nakamura (rumi.nakamura@oeaw.ac.at)

Key Points:

- Evolution of localized fast flows and dipolarization front is obtained from multi-scale multi-point observations in near-Earth magnetotail
- Current sheet thinning accompanied by intense field-aligned currents is detected following the passage of the dipolarization front
- From signatures of adiabatic electron acceleration it is confirmed that the same flow front was detected by the multi-point measurements.

**Abstract**

We report a unique conjugate observation of fast flows and associated current sheet disturbances in the near-Earth magnetotail by MMS (Magnetospheric Multiscale) and Cluster preceding a positive bay onset of a small substorm at ~14:10 UT, Sep. 8, 2018. MMS and Cluster were located both at X ~-14 $R_E$. A dipolarization front (DF) of a localized fast flow was detected by Cluster and MMS, separated in the dawn-dusk direction by ~4 $R_E$, almost simultaneously. Adiabatic electron acceleration signatures revealed from comparison of the energy spectra confirm that both spacecraft encounter the same DF. We analyzed the change in the current sheet structure based on multi-scale multi-point data analysis. The current sheet thickened during the passage of DF, yet, temporally thinned subsequently associated with another flow enhancement centered more on the dawnward side of the initial flow. MMS and Cluster observed intense perpendicular and parallel current in the off-equatorial region mainly during this interval of the current sheet thinning. Maximum field-aligned currents both at MMS and Cluster are directed tailward. Detailed analysis of MMS data showed that the intense field-aligned currents consisted of multiple small-scale intense current layers accompanied by enhanced Hall-currents in the dawn-dusk flow-shear region. We suggest that the current sheet thinning is related to the flow bouncing process and/or to the expansion/activation of reconnection. Based on these mesoscale and small-scale multipoint observations, 3D evolution of the flow and current-sheet disturbances was inferred preceding the development of a substorm current wedge.

**Plain Language Summary**

This is optional but will help expand the reach of your paper. Information on writing a good plain language summary is available here.

# 1 Introduction

In the near-Earth magnetotail, a major energy conversion process takes place associated with dipolarization and localized fast flows, called bursty bulk flows (BBFs) containing minute-scale flow bursts (FB)s (e.g., Baumjohann et al., 1989; Angelopoulos et al., 1994) with a mesoscale (few $R_E$: Earth radii) dawn-dusk size (Nakamura et al., 2004) that can ultimately affect the large-scale magnetotail dynamics (Sergeev et al., 2012, and reference therein). These localized BBFs contain sharp enhancements in Bz (south-to-north component of the magnetic field) called dipolarization fronts (DF) (Nakamura et al., 2002a). The thin DF, which has a thickness of 1-3 ion- inertial length (and/or gyroradii) (Runov et al., 2009; Schmid et al., 2011), is the leading boundary of a dipolarizing flux bundle (DFB) (Liu et al., 2013), which is a strong magnetic field region, associated with significant Earthward transport of magnetic flux.

Azimuthally localized BBFs and DFs have been considered to be associated with the transient /localized reconnection and different interaction processes between a localized fast flow and the ambient plasmas. Evolution of localized reconnection in a 1D Harris type current sheet has been simulated by 3D Hall-MHD simulation (Shay et al., 2003; Nakamura, T., et., 2012) and 3D PIC simulation (Liu et al., 2019) and the effect of a localized reconnection region on the evolution of the reconnection jets was shown. PIC simulations with normal component of the magnetic field to the initial current sheet, as expected in the near-Earth tail current sheet, reported DF like feature that leads to an onset of new reconnection at its trailing thin current sheet (Pritchett and Coronitti, 2011, 2013; Pritchett, 2013; Sitnov et al., 2013, 2017, 2021). Azimuthally localized reconnection jets interacting with dipole field region have been studied based on regional MHD simulation of magnetic reconnection (e.g., Birn et. 2011) as well as global MHD simulation (Merkin et al. 2019) to characterize the large-scale flow braking dynamics with flow bouncing and formation of vortical flow pattern.

A BBF streaming Earthward toward near-Earth dipolar region creates shear in flow and field and generates current systems with field-aligned currents that connect with the ionosphere. Such current systems created by the individual local DFB is shown to have the same polarity as the large scale substorm current wedge (Liu et al., 2013, 2015; Kepko et al., 2015, and references therein). Multi-point observations by the Time History of Events and Macroscale Interactions during Substorms (THEMIS) mission, which were separated by a fraction of $R_E$ to several of $R_E$, succeeded to monitor localized vortex signatures associated with flow braking of BBFs (Keiling et al., 2009; Keika et al.,2009; Panov et al., 2010ab). This flow pattern was predicted from MHD simulations (Birn et al., 2004; Birn et al., 2011) to produce the BBF associated field-aligned current system. PIC simulation of smaller scale flows such as interchange-generated finger structures (Pritchett and Coroniti, 2013) showed diversion of the electron currents that produce a field-aligned current pattern with same sense to the current wedge type currents around BBF. THEMIS observations compared with simulation (Panov and Pritchett, 2018; Panov et al., 2020) showed supporting evideneee that such kinetic ballooning/interchange instability is the source of DF and auroral streamers (Pritchett et al., 2014). Field-aligned currents connected to the Hall current in the diffusion region as shown in simulation studies (e.g., Ma and Lee, 2001) was also reported in the midtail observation associated with fast flows (Ueno et al., 2003) and in the flow braking region (Nakamura et al., 2018).

In this study we report a unique conjugate observation of a localized fast flow associated with a dipolarization front and current sheet disturbances observed almost simultaneously by MMS (Magnetospheric Multiscale) (Burch et al., 2016) and Cluster (Escoubet et al., 2001) preceding a small substorm (AL~ -200 nT) from 1400 UT, 8 September, 2018. Cluster and MMS were separated by about 4 $R_E$ mainly in the dawn-dusk direction in the near-Earth magnetotail at a downtail distance of about 14 $R_E$. The interspacecraft separation among the four MMS was about several 10 km, which was comparable to the two Cluster spacecraft separation distance. This configuration allows us to perform unique multi-scale gradient measurements of the dipolarizaton front and flow bursts together with the background magnetotail current sheet evolution.

The Cluster data used in this study are from the Cluster ion spectrometry (CIS) experiment (Rème et al., 2001), the Plasma Electron And Current Experiment (PEACE) (Johnstone et al., 1997), the FluxGate Magnetometer (FGM) experiment (Balogh et al., 2001), and the Electric Field and Wave (EFW) instrument (Gustafsson et al., 2001). The MMS data used in this study are obtained from the Fast Plasma Instruments (FPI) (Pollock et al., 2016), the Flux Gate Magnetometer (FGM) (Russell et al., 2016), the Search Coil Magnetometer (SCM) (Le Contel et al., 2016), and of the electric field obtained by the double probe instrument (EDP) (Ergun et al., 2016; Lindqvist et al., 2016).

## 2 Context of the event

### 2.1. Overview

A transient electroject intensifications took place at ~14:22 UT during a weakly disturbed interval on September 8, 2018 (Figure 1a). The midlatitude magnetograms showed that enhancement of Pi2 activity started around 14:01 UT (Figure 1d), associated with a positive bay enhancement (Figure 1b) with $B_Y$ deflection (Figure 1c). The latter started to develop gradually, but then was followed by a clearer enhancement after 14:11 UT. These midlatitude magnetic $B_X$ and $B_Y$ changes suggest the development of a relatively weak but distinct substorm current wedge (SCW) (McPherron et al., 1973). Its local time distribution was inferred from the model of Sergeev et al. (2011) using the mid/low-latitude ground-based magnetic field disturbances (See supplement material, the obtained local time of the wedge current system and the total current are shown in Figure S1). The fast flow event discussed in this study took place between 14:01 and 14:11 UT preceding the enhancement of the electrojet, but associated with the midlatitude magnetic activity, indicating the developing stage of a SCW.

Both MMS and Cluster were located in the premidnig sector near the center of the nominal current sheet (Figures 1e, h): MMS was located at (-13.8 , 6.1, 0.6) $R_E$, whereas Cluster was located at (-15.0 , 2.2, 3.2) $R_E$ in the Geomagnetic Solar Magnetospheric (GSM) coordinate system. All the figures in this paper are shown in the GSM coordinate system, which describes well the current sheet orientation of this event as will be discussed more detail in section 2.2. The relative distance was $\Delta R_{(Cluster-MMS)}$ = (-1.2, -4.0, 2.6) $R_E$. The interspacecraft distance of MMS was about 45 km (Figures 1f and 1i). Due to a solar eclipse maneuver, only Cluster 3 and 4 data were available for this event, which were separated about 15 km mainly in Z direction (Figures 1g and 1j). Field-lines that are crossing MMS and Cluster are drawn in in Figures 1e and 1h using Tsyganeko 15 magnetic field model (Tsyganenko and Andreeva, 2015). The local time of the mid/low latitude stations used in obtaining the location of SCW are given in Figure 1e.

According to the SCW model (Figure S1), both MMS and Cluster were located close to duskside of SCW after 14:13 UT expanding eastward. A clear large scale SCW, however, was not developed during the time interval of MMS and Cluster flow event so that modeling was not possible. Yet, as can be seen Figure 1b, a gradual enhancement of positive bay started already after 14:01 UT as a dipolar pattern in deflection ($B_Y$), that is negative $B_Y$ in the dawn side stations (red and blue traces) and positive in the dusk side stations (black and green traces) expected for SCW, was gradually growing and became more visible after 14:11 UT in Figure 1d. The local time distribution of the magnetic disturbance at 14:12 UT (Figure S2a) shows that a more distinct SCW-like disturbance (at least expected for dusk-side part) has developed at the premidnight region, which then developed at a later time toward a larger SCW disturbance detected over a wide local time area as shown in Figure S2b for 14:25 UT.

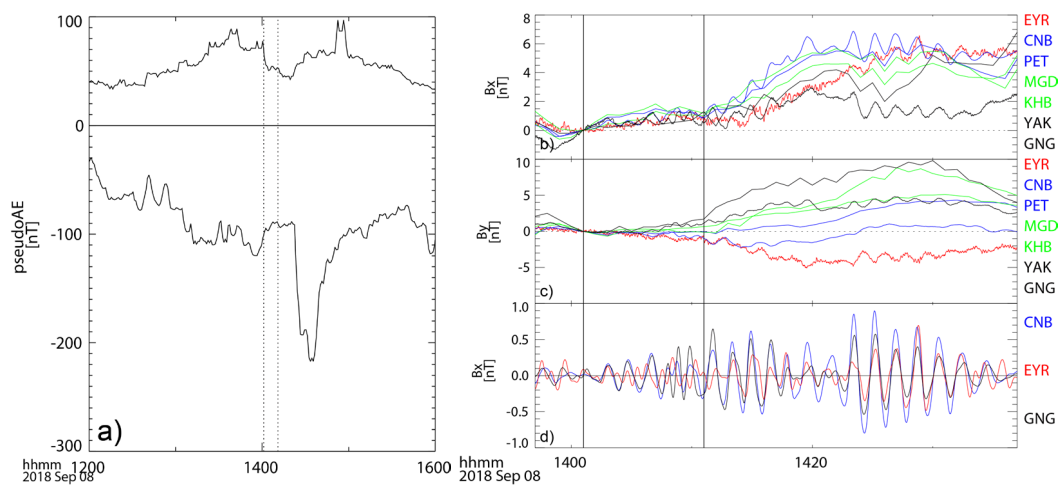

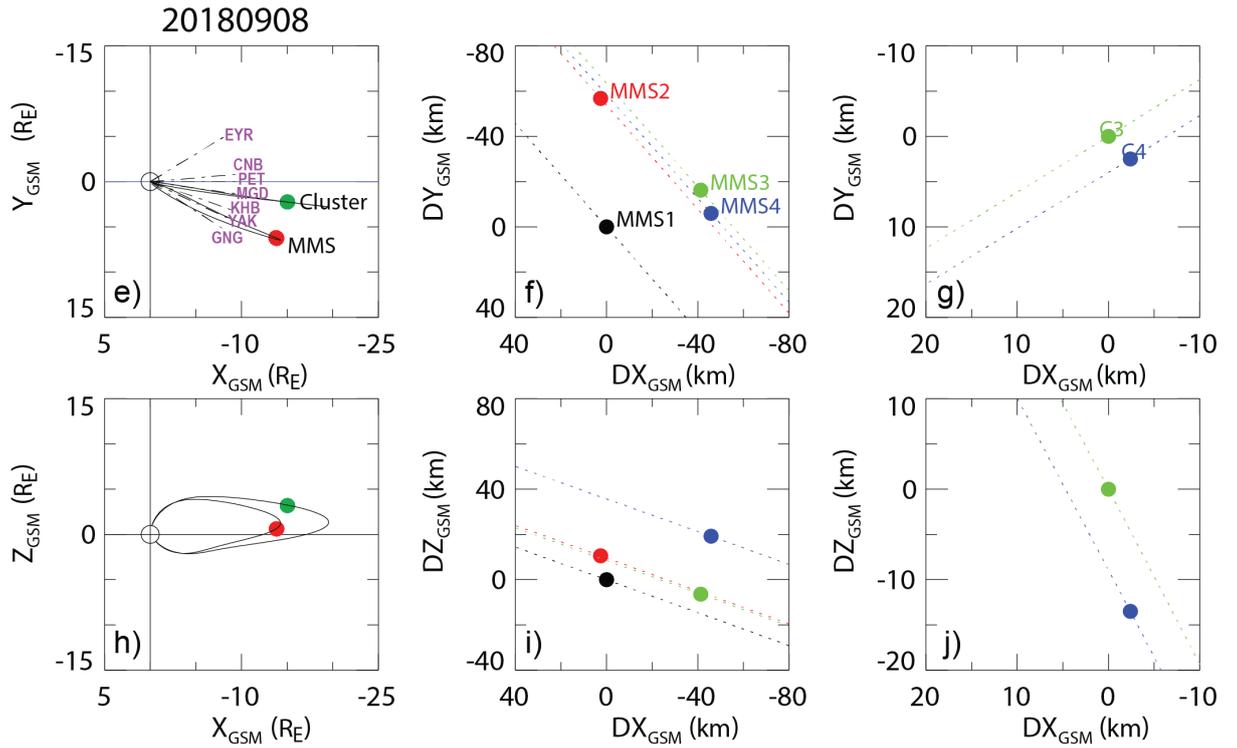

Figure 1. Ground-based magnetograms and location of MMS and Cluster spacecraft. (a) Pseudo AE, (b) northward (Bx) and (c) eastward (By) component of mid-latitude magnetogram. While data from the southern hemisphere stations, are 4-s averaged data, all the other plots of the ground-based magnetogram are showing 1-min resolution data. (d) Filtered Bx component for the three southern hemisphere stations for frequency between 7-25 mHz. The vertical lines indicate 14:02 UT and 14:12 UT. Location of MMS and Cluster in (e) GSM *X-Y* and (f) *X-Z* plane with model field lines. Magnetic local time of mid-latitude ground based station 14:00 UT are indicated by dashed lines in (e). Location of the four MMS spacecraft relative to MMS1 in (f) *X-Y* plane and (i) *X-Z* plane. Relative location of C3 and C4 with in (g) *X-Y* plane and (j) *X-Z* plane. The dotted lines in (f-g) and (i-j) present the spacecraft orbit. Field-line crossing MMS and Cluster using Tsyganeko 15 magnetic field model in (e) and (h).

An overview of the Cluster and MMS observations of the fast flow event is shown in Figure 2. Both Cluster and MMS observed an enhanced flow interval starting around 14:01 UT with enhancement in $B_z$ nearly simultaneously, indicating a dipolarization front. While Cluster observed high-speed Earthward flow up to about 1000 km/s near the equator and northern hemisphere, where $B_X$ is positive, MMS observed Earthward flow with a speed of about 200 km/s at the southern hemisphere at off-equatorial plasma sheet, as can be seen in the negative $B_X$ profile. Cluster observed another enhancement of Earthward flow with a speed up to 1500 km/s starting after 14:05 UT, when MMS observed enhancement of duskward and then mainly tailward flow. At 14:11 UT, the fast flow event at Cluster ceased and the Earthward flow changed toward tailward, while MMS observed reversal from tailward to Earthward flow.

These near simultaneous changes of the flow between Cluster and MMS suggest that both spacecraft observed the same BBF-associated disturbances at different location at the beginning of the event. This conjecture is supported by the results of detailed analysis of magnetic field and plasma signatures described in the following sections. From the larger speed of the flows it is suggested that Cluster was located closer to the center of BBF, while MMS was located at the edge of BBF in terms of local time. The flow is expected to be localized and temporally changing, as inferred from the changes in all components of the flow as well as the field also in the subsequent flow enhancements. There is a flow reversal around 14:11 UT for which the sense of change in direction is opposite between MMS and Cluster, i.e. Earthward to tailward at Cluster and tailward to Earthward at MMS. Such changes in the vortex direction identified by multi-spacecraft separated in dawn-dusk direction has been reported also in previous observations by THEMIS (Panov et al, 2010) and by Cluster ( Nakamura et al., 2013). This reverse in the flow vortex direction suggest a flow bouncing signature of a localized flow, which was also seen in the MHD simulations (Birn et al., 2011; Merkin et al.,2019) at flow braking region.

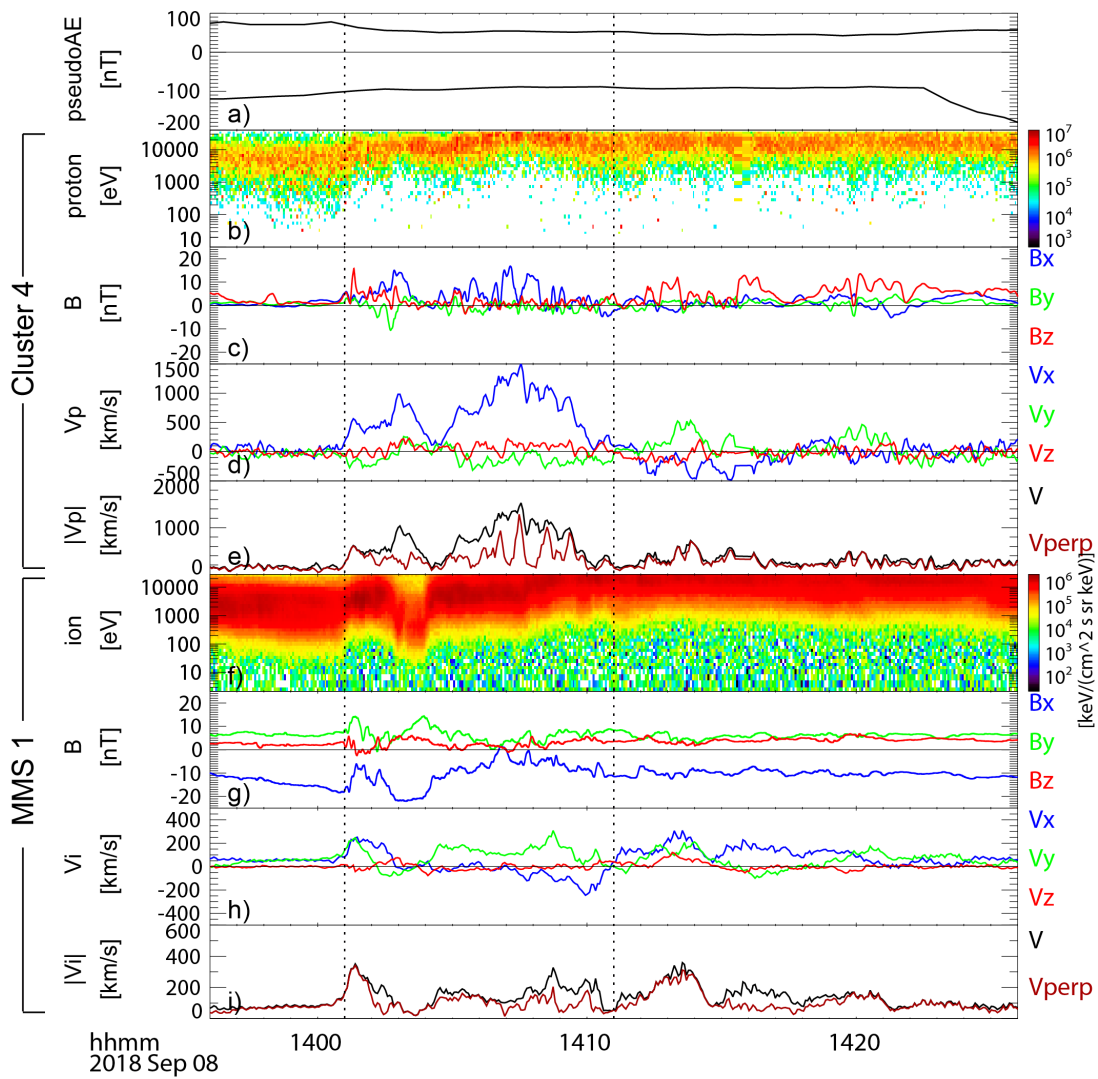

Figure 2. Overview of Cluster and MMS observation for September 8, 2018, 13:50-14:30 UT. (a) Pseudo AE, (b) proton energy spectra, (c) magnetic field, (d) proton flow, and (e) magnitude of proton flow from Cluster 4; (f) ion energy spectra, (g) magnetic field, (h) ion flow, and (i) magnitude of ion flow from MMS 1. The *X*, *Y*, and *Z* components in GSM coordinates are shown in blue, green, and red colors in panels (c), (d), (g), and (h). Magnitudes of the total flow and perpendicular flow are plotted in black and brown in panels (e) and (i). The vertical dotted lines are the start of Pi2 (14:01 UT) and positive bay (14:11 UT), respectively

## 2.2 Current sheet change and spacecraft position

The magnetic field changes at both MMS and Cluster during the flow interval are not only in the normal component ($B_Z$), but significant changes are observed also in the $B_X$ component as can be seen in Figures 3a and 3b. That is, the structure of the current sheet and/or the position of the spacecraft relative to the background current sheet are changing during this interval. Nonetheless, the overall temporal evolution of the total tail current, that can be deduced from the estimated lobe field value, $B_0$, is only gradually increasing as can be seen in Figure 3c. This gradual evolution is expected from the magnetotail loading phase preceding the weak substorm at 14:22 UT. The lobe field value, $B_0 = (2\mu_0 P_{tot})^{1/2}$, shown in Figure 3c is obtained from the total pressure, $P_{tot} = P_p + P_B = P_p + (B_X^2 + B_Y^2)/2\mu_0$, where $P_p$ and $P_B$ are the plasma pressure and the magnetic field taking into account only the horizontal component, respectively. The average $B_0$ during this interval is 31nT and 33nT for MMS and Cluster, respectively.

In order to obtain the context of the current sheet during the flow interval, we analyze the average magnetic field differences between the MMS and Cluster using a simple current sheet model. To examine the change in the background current sheet, we use 20s boxcar averaged magnetic field data to eliminate the small/short-time scale local disturbances for the magnetic field. Furthermore, although both MMS and Cluster are multipoint missions in themselves, we first consider Cluster and MMS as two-point measurements with respect to the large-scale (>4 $R_E$) background current sheet. Based on the neutral sheet model by Tsyganenko et al. (2015) that takes into account the effect of the dipole tilt angle (12.7°), IMF $B_Z \sim$ -1.1 nT, IMF $B_Y \sim$1.3 nT and the dynamic pressure of 2.3 nPa as input parameters from 2018/09/08 14:00 UT, the hinging distance was estimated to be 12.5~12.7 $R_E$ at MMS and Cluster local time. That is, both spacecraft can be considered well tailward of the hinging distance. The normal direction of the neutral sheet deduced from the change in the neutral sheet position in the model within ±0.5 $R_E$ around Cluster and MMS location was close to the $Z_{GSM}$ axis with only a slight tilt (8°) toward +Y axis for MMS and with less than 3° tilt for Cluster. In addition to the effect from the large-scale solar wind-magnetospheric interaction modeled by Tsyganenko et al. (2015), large-scale current sheet waves (e.g., Sergeev et al., 2003; Vasko et al., 2014) may further change the tilt of the current sheet in the *Y-Z* plane. To evaluate these effects in our observation we examined the averaged MMS current density (curlometer) between 14:00 and 14:10 UT. Note that these local gradient observations can contain fluctuations of different scales including transient small-scale currents (discussed more detail in section 3.2) in addition to the changes in the large-scale current sheet. From the average and standard deviation of the perpendicular current, $j_Y$ $_{GSM}$=2.2 ±4.5 nA/m$^2$ and $j_Z$ $_{GSM}$ = -0.1 ±2.9 nA/m$^2$ we obtain that the perpendicular current was slightly tilted (2.5° ±7.0°) toward –$Z$ $_{GSM}$ axis in the *Y-Z* plane. The current density direction is therefore consistent with the current sheet orientation obtained from the Tsyganenko model on average and does not significantly deviate from the $Y_{GSM}$ axis even taking into account the fluctuations. We therefore assume a planar current sheet in the GSM coordinate system and use a 1D Harris current sheet (Harris, 1962) configuration, $B_X = B_0 \tanh((Z-Z_0)/D)$ where $Z_0$ is the current sheet center, D is the half thickness of the current sheet, and $B_0$ corresponds to the field outside the current sheet. It should be noted that the results discussed below will not change even if we took into account the effect of the tilt of the current sheet discussed above. We use the $B_0$ profile from MMS data (red curve in Figure 3c for the model) also at Cluster location by correcting the small difference (5%) in magnitude as shown in the blue line in

Figure 3c. Using this simplified current sheet model we infer the current density profile and relative location of the two spacecraft to the current sheet as shown in Figure 3d. It can be seen that following the encounter of the dipolarization at 14:01UT the current sheet becomes thick. Interestingly this current sheet thickening at 14:01:20-14:01:50 UT is followed by a thinning of the current sheet with enhanced current density centered around 14:02:20-14:03:40 UT. This change can also be seen in the modeled current density at the center of the current sheet $J_0 = B_0/D$, as well as in the current density, $J_{CL}$ ($J_{MMS}$), at Cluster (MMS) location (Figure 3e).

The magnitude of the current density is also determined from the local magnetic field gradient among the four MMS and two Cluster spacecraft and shown in Figure 3f. Using the typical parameters around the current sheet thinning at 14:02:30 UT ($n$~0.6 /cc, $B$~15 nT, $T_i$~ 2 keV at MMS and $T_i$ ~ 3 keV at Cluster, where $n$ is the density and $T_i$ is the ion/proton temperature at MMS/Cluster), the ion inertia length, $d_i$, is estimated to be $d_i$ ~ 290 km and ion gyro radius, $\rho_i$, at Cluste to be $\rho_i$ ~ 380 km and at MMS to be $\rho_i$~ 310 km. Hence the estimation of the current density assuming a linear gradient using Cluster or MMS will have sufficient resolution to obtain structure down to sub-ion scales. The temporal changes of the current density deduced from the large-scale current sheet model (Figure 3e) can be seen also in the local current density (Figure 3f). That is, both current density values show the decrease of the current density associated with the dipolarization front followed by intensification of the current density. It should be also noted that Cluster has only data from two spacecraft due to eclipse maneuver. The two Cluster spacecraft separation across the current sheet as well as the spatial scale of the MMS tetrahedron is within several tens km (see Figure 1) and comparable current density values are obtained at Cluster and MMS. Yet, these current density values shown in Figure 3f are about 5 times larger than those determined from the large-scale gradient between Cluster and MMS. This is possible due to the simplified current profile model as well as the large distance (>2 $R_E$) between Cluster and MMS across the current sheet, which would not allow to model a thin embedded current sheet expected to be observed at growth phase in near-tail region (Petrukovich et al., 2011, 2015; Artemeyev et al., 2021). Furthermore, it is apparent that the local current density in Figure 3f from both Cluster and MMS have a shorter time-scale enhancement at the beginning of the current sheet thinning. This peak in the current density was observed both at Cluster and MMS when they are located in off-equatorial region (Figure 3d). The subsequent decrease in current density corresponds to the time when Cluster approaches equator while MMS went out from the current sheet center. The profile therefore suggests a current sheet with off-equatorial peaks, or bifurcated current sheet, since both spacecraft observed these off-equatorial peaks around the same time. Previous Cluster observations reported evidences of bifurcation during a fast flow event associated with current sheet thinning (Nakamura et al., 2002b) as well as in a number of quiet thin current sheet crossings (Runov et al., 2006). Comparison between the large-scale gradient and small-scale gradient measurement therefore allows to determine also the small-scale structures within the thinning current sheet. In the following we examine in more detail the plasma and field disturbances during the dipolarization front and subsequent period where current sheet is dynamically changing between 14:01 UT and 14:04 UT.

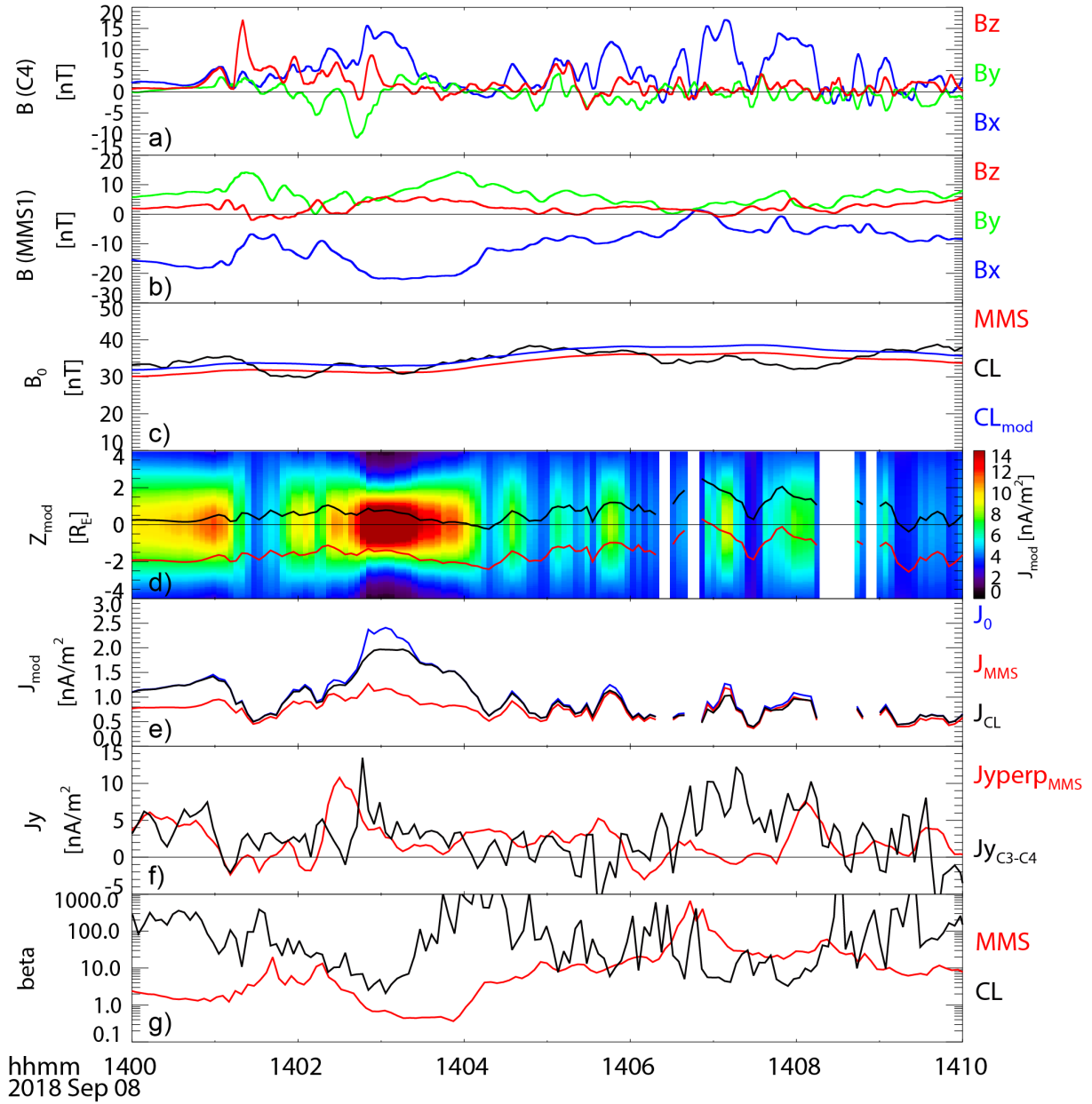

Figure 3. Change in the current sheet configuration and spacecraft location relative to the current sheet. Magnetic field observed at (a) Cluster 4 and (b) MMS1. (c) Magnetic field outside the current sheet, $B_0$, estimated from the total pressure at MMS1 (red) and at Cluster 4 (black) and the fitted $B_0$ at Cluster using MMS1 data (blue). (d) Current density distribution of the modeled current sheet and location of the two spacecraft, MMS (red) and Cluster (black), within the model current sheet. (e) Current density distribution obtained from the current sheet model at equator (blue), at location of MMS (red) and Cluster (black) relative to the current sheet center, (f) dawn-dusk component of the current density obtained from curlometer method for perpendicular component at MMS (red) and that obtained from C3 and C4 (black), (g) Plasma beta for MMS (red) and Cluster (black).

## 3. Evolution of flow and field associated with the dipolarization front

We examine the structure and evolution of BBF and dipolarization front in detail by using the position of the MMS and Cluster spacecraft relative to the current sheet.

### 3.1. Dipolarization front and flow shear

Figure 4 shows Cluster and MMS observations during the encounter of the fast flow and dipolarization front. As discussed before both MMS and Cluster observed $B_Z$ enhancement associated with the Earthward flow nearly simultaneously: MMS at 14:01:02 UT and Cluster at 14:01:12 UT (shown as vertical lines in Figures 4b-g). Since MMS was located away from the current, this time difference is consistent with the direction of Earthward/Equatorward motion of a flux tube so that MMS will first encounter the dipolarized flux bundle (DFB). Hence it is most likely that Cluster and MMS encountered the same flow associated DFB. The flows perpendicular to the magnetic field (Figures 4c, f, h) in the *X-Y* plane around the dipolarization front indicate clock-wise/counter clock-wise flow rotation viewed from north for Cluster/MMS, suggesting that Cluster/MMS was located at the dawnside/duskside part inside a flow channel.

We examine magnetic disturbance of the front using the methods introduced by Liu et al. [2013] for which the orientation of the dipolarization front current sheet (DFCS) is determined based on three different methods by using the results from the minimum variance analysis and the magnetic fields inside (immediately behind) and outside (immediately ahead) of the dipolarization front, **B**$_{in}$ and **B**$_{out}$. Following the definition of DF event by Liu et al. [2013], we first use 3s averaged data to determine the start time of DF, $t_0$, which is the first time that meets the criteria $dB_Z/dt$>0.5 nT/s applied on the three-point running average data. **B**$_{in}$ and **B**$_{out}$ are the magnetic field from the following peak and the preceding dip values of the bipolar $B_Z$ change closet to $t_0$ and given in Table 1. The normal direction, **n**, to the DFCS is then determined in the following three different ways to examine the spatial form of DFCS: (1) minimum variance direction, **N**, (2) **B**$_{in}$ × **B**$_{out}$ and (3) **B**$_{in}$ × **L**. Here **N** and **L** are the minimum and maximum variance directions, respectively, obtained from the minimum variance analysis around the DFs performed using burst mode magnetic field data for the time intervals given in the first row in Table 1. The obtained normal directions are then expressed by the elevation angle, $\theta_{\mathbf{n}\text{-}XY}=\sin^{-1}n_Z$, which is the angle between **n** and the background tail current sheet plane (*X-Y* plane), and the azimuth angle, $\alpha_\mathbf{n}=\sin^{-1}(|n_Y|)$. Liu et al. [2013] obtained a distinct pattern of these angles as well as the angle between **B**$_{in}$ and **L**, $\theta_{\mathbf{B}_{in}\text{-}L}$, depending on the spacecraft location relative to the DFCB, which is expected to have a convex shape in equatorial cross section and dipolarized field line shape in meridional cross section. This relationship enables us to identify the relative azimuthal location of DF (duskside or dawnside) and hemisphere (north or south).The ratio of the average *X* component of the magnetic field to the lobe field value during quiet time, $B_{qx}/B_{lobe,q}$, is used as an indicator of the position relative to the equator within the plasma sheet. Here "q" indicates the quiet time average for time interval before the DF event, for which we used the time between 13:59 and 14:00 UT. The lobe field, $B_{lobe}$ is the same as $B_0$.

Table 1 shows also these angles obtained for the observed DF. All three methods obtain positive (negative) $\theta_{\mathbf{n}\text{-}XY}$ >0 (<0) for MMS (Cluster) staying in the southern (northern) hemisphere, where negative $B_{qx}/B_{lobe,q}$ <0 (>0) is expected for the front shape of a flux-tube (DFB). At MMS location positive $\alpha_\mathbf{n}$ > 30° was obtained, indicating that the spacecraft was close to the DF duskside flank and expected therefore larger $\theta_{\mathbf{B}_{in}\text{-}L}$ compared to Cluster, for which -30°< $\alpha_\mathbf{n}$ < 0°, indicated being in the morning side, but closer to DF head line as well as closer to the equator, with small $|B_{qx}/B_{lobe,q}|$, therefore expecting small $\theta_{Bin-L}$. That is, the dipolarization front observed by Cluster is oriented closer to the *Y-Z* plane than for MMS with small tilt consistent with a localized dipolarization front direction at northern hemisphere and dawnside of the flow center. The dipolarization front from MMS, on the other hand, is more tilted both in the *X-Y* and *X-Z* plane suggesting southern hemisphere and dusk side of the flow center. The normal vector from minimum variance analysis is shown as blue solid line and the cross section of DF and the *X-Y* (*Y-Z*) plane is shown as blue dotted line in Figures 4i (4j). These relationships as well as the obtained changes in the current sheet thickness behind the dipolarization front suggest that MMS and Cluster encounter a localized DFB around 14:01 UT with consistent magnetic field disturbance followed by the local current sheet thickening as obtained by Liu et al. (2013). This is also consistent with the two spacecraft analysis of the current sheet obtained shown in Figure 3. As expected, there is flow shear observed around the DF at MMS1 toward dusk while for Cluster a dawnward deflection as can be seen in Figure 4h.

**Table 1** Orientation of the dipolarization front observed by MMS 1 and Cluster 4.

|  | MMS1 | Cluster4 |
|---|---|---|
| Minvar Coordinate **L**, **M**, **N** (GSM) and eigenvalue ratios: $e_L/e_M$ and $e_M/e_N$ | (14:01:00-14:01:12 UT) **L**=(-0.38, -0.06, 0.92) **M**=(-0.54, 0.82, -0.17) **N**=(0.75, 0.57, 0.35) $e_L/e_M = 3$, $e_M/e_N = 13$ | (14:01:00-14:01:20 UT) **L**=(0.21, 0.06, 0.98) **M**=(0.39, 0.91, -0.14) **N**= (0.90, -0.41, -0.17) $e_L/e_M = 8$, $e_M/e_N = 14$ |
| Dipolarization front ($t_0$) | 14:01:02 UT | 14:01:12 UT |
| **B**$_{in}$ (GSM) | (-18.5, 9.8, 5.5) [nT] | (4.9, 2.4, 22.0) [nT] |
| **B**$_{out}$ (GSM) | (-15.9, 8.8, 0.4) [nT] | (1.1, 1.5, 0.3) [nT] |
| ($\theta_{n-XY}$, $\alpha_n$) from minvar | (20°, 34°) | (-10°, -24°) |
| ($\theta_{n-XY}$, $\alpha_n$) from **B**$_{in}$ × **B**$_{out}$ | (5°, 60°) | (-7°, -35°) |
| ($\theta_{n-XY}$, $\alpha_n$) from **B**$_{in}$ × **L** | (16°, 55°) | (-11°, -35°) |
| $\theta_{Bin-L}$ | 43° | 2° |
| $B_{qx}/B_{lobe,q}$ | -0.24 | 0.01 |

*Note.* **B**$_{in}$ and **B**$_{out}$ are magnetic fields inside and outside (immediately ahead) of the dipolarization front. The orientation of the dipolarization front is expressed by the elevation angle, $\theta_{n-XY}$, and azimuth angle, $\alpha_n$. $\theta_{n-XY}$. $\theta_{Bin-L}$ is the angle between **B**$_{in}$ and **L** and $B_{qx}/B_{lobe,q}$ is the ratio between the quiet magnetic field to the lobe field. More detail methods to determine these angles and their relationships to the orientation of the dipolarization front are given in Liu et al. (2013).

There is another dawn-dusk flow shear apparent around 14:02 UT in Figure 4h, around the time of the drastic change in the current sheet configuration (Figure 4a). The perpendicular flow changes from duskward to dawnward at MMS, while for Cluster the flow changes from dawnward to duskward. Consistent with this flow pattern the electric field changes from southward to northward, as expected from $-\mathbf{V} \times \mathbf{B}$ electric field for spacecraft at south and duskside (MMS). Although the sense of the flow rotation viewed from north, i.e. clockwise (anticlockwise) for Cluster (MMS), are in the same sense as that at the dipolarization front around 14:01 UT, these flow shear is not necessary showing the equatorial flow pattern (i.e., flow pattern in *X-Y* plane) as describe below.

As shown in Figure 4a the flow shear after 14:02 UT is associate with the thinning of the current sheet as well as gradual northward motion of the current sheet relative to the spacecraft as can be also seen in the dashed line in Figure 4i. That is, MMS should be seeing more off equatorial signature due to thinning as well as outward motion relative to the current sheet center. An dawnward flow toward the center of BBF at the off equatorial region has been see in the 3D simulation (e.g. Birn et al.,2011) and was observed by MMS (Nakamura et al., 2018) at the dusk part of BBF as illustrated in the Figure 4j. The flow shear after 14:02 UT observed MMS is therefore likely due to such encounter of off equatorial inflow region of azimuthally localized BBF.

Cluster, on the other hand, approached closer to the equator of the thin current sheet during the enhanced duskward flow. This pattern cannot be explained if Cluster observed the dawnside part of a narrow flow channel as was the case during the encounter of the dipolarization front around 14:01 UT as illustrated in the mid panel of Figure 4j. Instead, it is more plausible to consider that Cluster was located also at the duskside of BBF as illustrated in the right panel of Figure 4j. Cluster observed another enhancement Earthward flow (~1000 km/s, mainly parallel flow) with Bz enhancement associated with the current sheet thinning at 14:02:30 UT. That is, the flow changes after 14:02 UT is more consistent when we interpret that both MMS and Cluster observed flow shear located at the duskside of an extended BBF in a thinner current sheet, which was moving northward. We cannot differentiate whether this change in the flow and current sheet configuration took place as part of the BBF flow braking processes or due to encounter of different BBF (from a more extended source region). Either way, this multipoint observations by MMS and Cluster suggests complex evolution of the localized flows.

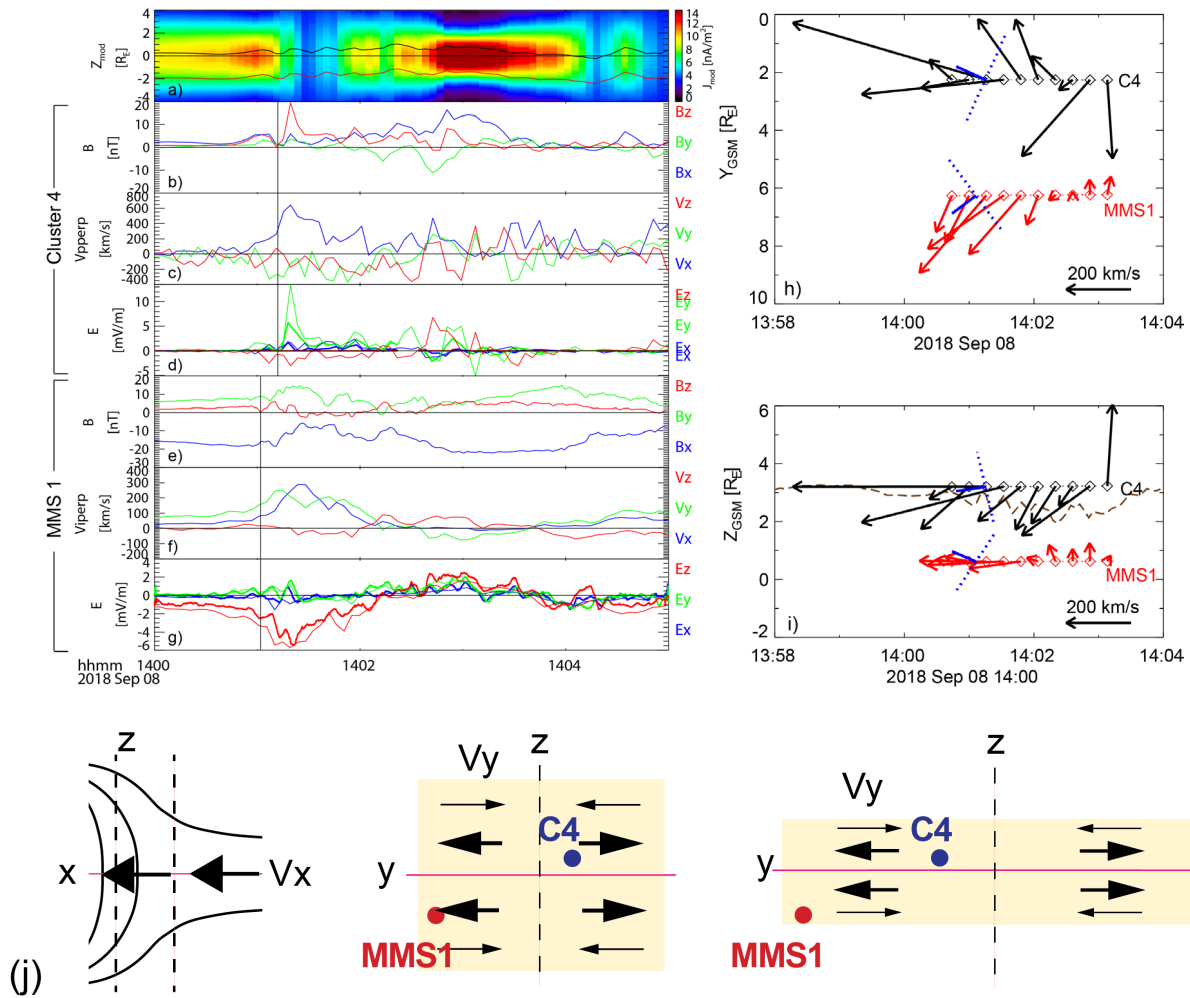

Figure 4. Cluster and MMS flow and fields observations between 14:01 UT and 14:05 UT. (a) Location of MMS (red) and Cluster (black) relative to the modeled current sheet center (same as Figure 3d); (b) magnetic field, (c) proton flow perpendicular to magnetic field, and (d) electric field observed by Cluster; (e) magnetic field, (f) ion flow perpendicular to magnetic field, and (g) electric field observed by MMS. Thin and thick lines in (d) and (g) are −**V**×**B** electric field and measured electric field, respectively. (For Cluster only *X* and *Y* components of the electric fiel are shown.) The *X*, *Y*, *Z* components are shown in blue, green, and red in panels (b-g). The vertical lines in (b-d) and (e-g) indicate the time of dipolarization front at Cluster (14:01:12 UT), and at MMS (14:01:02 UT), respectively. Temporal changes of flow vectors from C4 (black) and MMS1 (red) and dipolarization front (blue) in (h) the *X-Y* plane (upper panel) and in (i) the *X-Z* plane (lower panel). Each vector is 16 s average. The blue lines show the normal direction to the dipolarization front and the blue dotted lines indicate the projected direction of the dipolarization front from minimum variance analysis. The location of the current sheet from the model (Figure 4a) is shown as brown dashed curve. (j) Schematic representation of MMS1 and C4 location relative to the fast flow region during interval of thick current sheet (middle panel) and thin current sheet (right panel) in *Y-Z* plane.

## 3.2. Field-aligned currents

Figures 5 shows the changes in the field-aligned current and electron signatures associated with the flow event. The field-aligned current from Cluster shown in Figure 5b is calculated from 4-s particle moment data. The field-aligned current from MMS (Figure 5h) is estimated from the curlometer method and averaged to 4s resolution in order to compare with Cluster. The electron moment data shown in Figures 5l and 5m from MMS are from 4.5 s resolution fast mode data. The plotted energy spectra data are from the 30 ms burst mode data averaged over 40 point (1.2 s), which was available after 14:00:43 UT. As discussed before, the flow burst/DFB forms a field-aligned current pattern similar to a localized substorm current wedge (e.g., Liu et 2013; Kepko et al., 2015, and reference therein). From the orientation of the front and flow pattern shown in Figure 4 it is expected that Cluster (MMS), which was located at the dawnside (duskside) of the front, encounter downward (upward) field-aligned current toward (outward from) the ionosphere or positive (negative) $J_X$ and/or positive $J_Z$ behind the front. Figures 5b and 5h show that the observed field-aligned current direction during the encounter of the dipolarization front (start times are indicated by the solid lines) is consistent with such pattern. The electron spectra from both Cluster and MMS showed increase in energy (Figures 5c and 5i), density decrease (Figures 5g and 5m), and temperature increase (Figures 5f and 5l) as has been identified as typical signature of the dipolarization front (e.g., Runov et al., 2009).

The most intense field aligned current during this flow event, however, was observed later associated with the thinning of the current sheet (Figure 5a) at 14:02:46 UT by Cluster (indicated by dotted line in Figure 5b) and at 14:02:38 UT by MMS (Figure 5h). Here unlike the dipolarization front at 14:01 UT, Cluster detected also upward field-aligned current. Electron pitch-angle spectra plots from Cluster (Figures 5d and 5e) also support enhancements in predominantly parallel (0°) electron enhancement after ~14:02 UT. Note that Cluster likely encountered flow burst activity being centered further dawnward (Figure 4) so that Cluster also was located at the duskside of the flow burst at this time. That is, this later flow burst orientation expects that upward field aligned current (negative $J_X$) to be detected and is therefore consistent with the observed upward field-aligned current. The intense field-aligned current at Cluster is accompanied also by enhancement in Bz (Figure 4b), increase in energy of electron spectra (Figure 5c), density decrease, and temperature increase similar to the dipolarization front at 14:01 UT. The interpretation of the change in BBF location illustrated in Figure 4j is therefore supported also by the field-aligned current direction and electron spectra observed by Cluster.

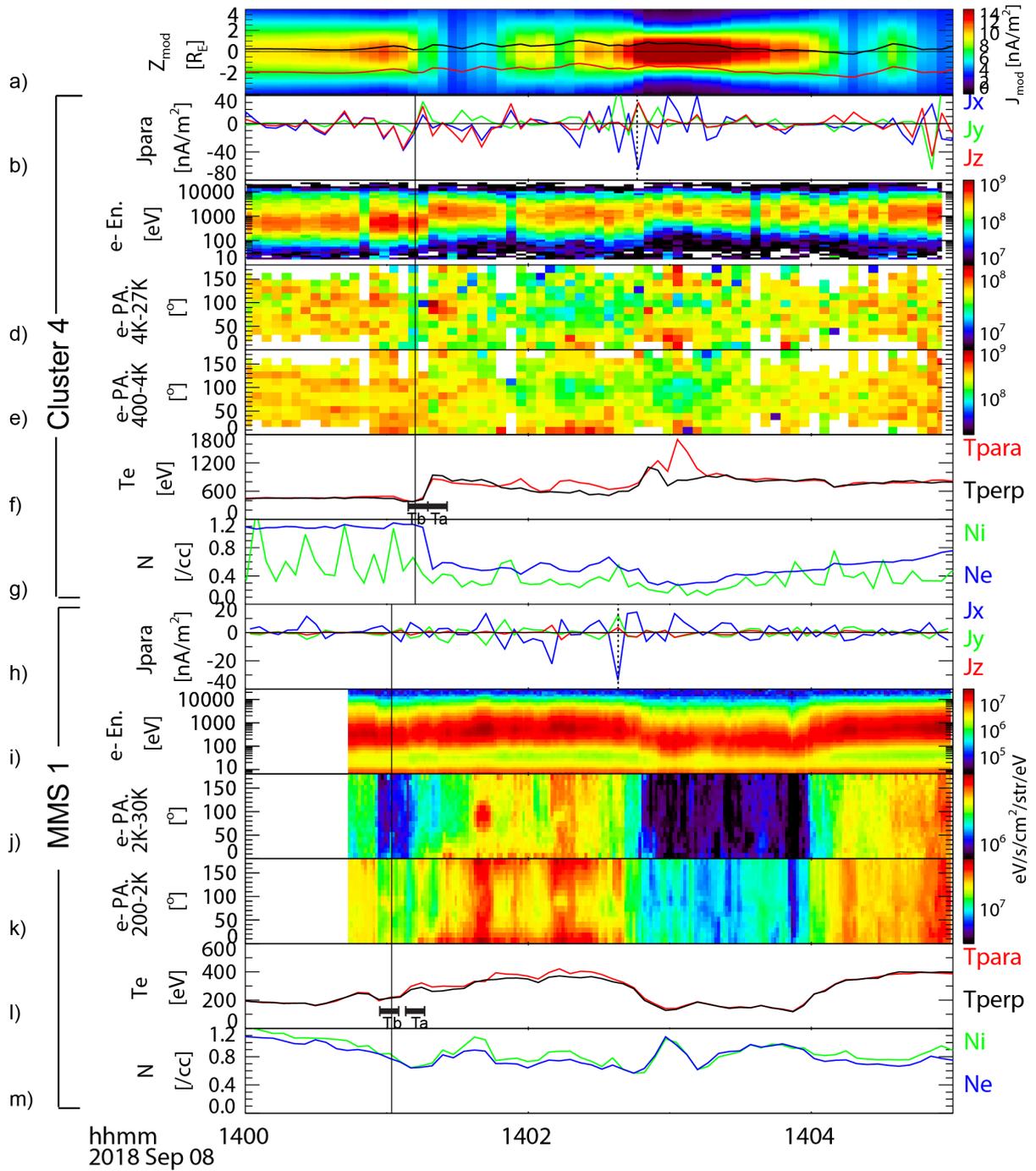

Figure 5. Field-aligned currents and electron data between 14:00 and 14:05 UT. (a) Location of MMS (red) and Cluster (black) relative to the current sheet (same as Figure 4a). (b) Field-aligned current calculated from particle moment; (c) electron omni-directional energy spectra and pitch-angle distribution for energy range (d) between 4 keV and 27 keV and (e) between 400 eV and 4 keV; (f) electron temperature and (g) density from Cluster 4. (h) Field-aligned current obtained from curlometer method, (i) electron omni-directional energy spectra and pitch-angle distribution for energy range (j) between 2 keV and 30 keV and (k) between 200 eV and 2 keV; (l) electron temperature and (m) density from MMS1. In (b) and (h), *X, Y,* and *Z* components of the field-aligned currents are given in blue, green, and red, respectively. Parallel (perpendicular) temperature is shown in red (black) curve in (f) and (l). Temperature from ions (electrons) are given by green (blue) in (g) and (h). The vertical solid lines show the DF time for MMS1 and Cluster 4 (see Figure 4). The vertical dotted lines in (b) and (h) indicate the times of the most intense field-aligned current at Cluster (14:02:46 UT) and (MMS 14:02:38 UT), respectively. The two thick black bars in (f) and (l) indicate the time intervals used for analysis of change in the particle distribution (see for details in Section 3.3.).

MMS also observed an intense upward field-aligned current associated with the current sheet thinning accompanied by enhanced anti-parallel (180°) field-aligned electrons (Figures 5j and 5k). Yet, the field-aligned current event is followed rather by decrease in overall energy in plasma consistent with MMS exiting from the center of the current sheet as discussed before. That is, the change in MMS reflects more a vertical structure of the thinning current sheet, including the flow shear in $V_Y$ and $E_Z$ reversal as discussed before. Figure 6 shows more detail of the field-aligned current and $E_Z$ at MMS using the high-resolution (burst mode) data. Figure 6b shows more structured parallel current in the high-resolution data. There is overall good agreement between the current obtained from the curlometer (blue trace) and that from the particle current (green), which is mainly carried by electrons. As was also shown in Figure 5, the strongest field-aligned current takes place during the thinning of the current sheet corresponding to the flow shear from duskward to dawnward, corresponding to the reversal of the *Z* component of the convection electric field, $-\mathbf{V}_i \times \mathbf{B}$, as shown in Figure 6c.

Figures 6c and 6d show how the different non-ideal MHD terms will contribute to the observed electric field. It can be clearly seen that the interval of the enhanced field-aligned currents is not only the time of the bulk flow reversal, but also related to the time when the non-ideal MHD terms, in particular the Hall-term $\mathbf{J}_{b,p} \times \mathbf{B}$, obtained from particle ($\mathbf{J}_p$) and magnetic field ($\mathbf{J}_b$) measurements, is significantly enhanced (Figure 6d), possibly then forming strong field-aligned current as closure current. The dominance of the Hall-term and small-scale process can be only seen during the time of strong field-aligned current, while at the beginning of the flow observation, when MMS was closer to the equator, the convection component ($-V_i \times B$) dominates and the intense field-aligned current disturbances were not detected.

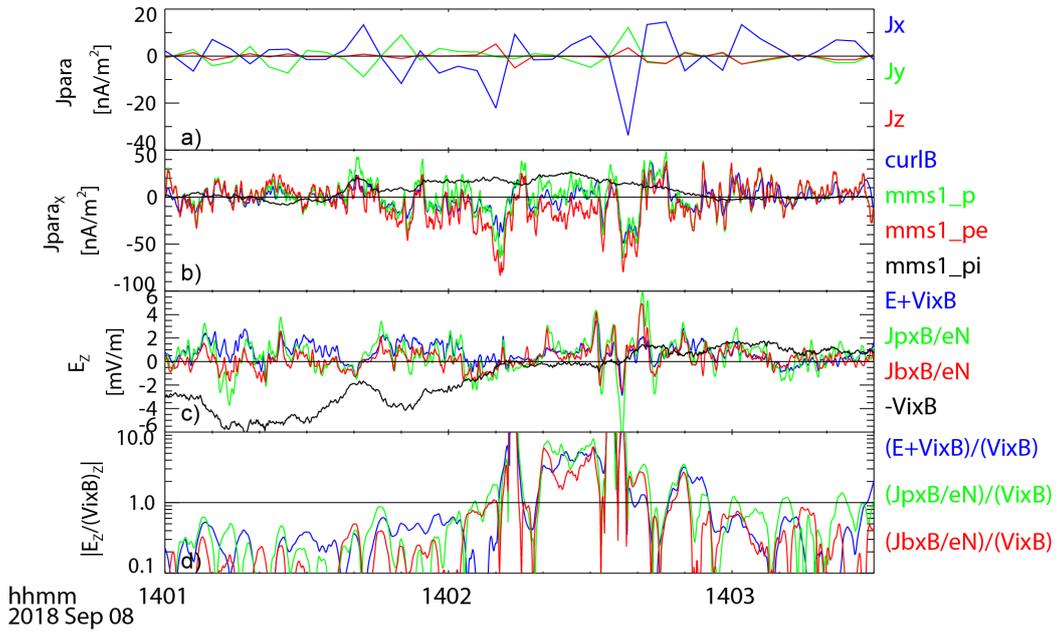

Figure 6. Detailed characteristics of field-aligned current and electric field from MMS1 high-resolution data between 14:01:00 and 14:03:30 UT. (a) 4-s averaged parallel current density obtained from curlometer; (b) Parallel current density from the curlometer method (blue), currents calculated from ions (black) and electrons (red), and total of ions and electrons (green) from MMS. (c) Z components of the perpendicular electric field: $E + V_i \times B$ (blue), where E is the measured electric field double probe instrument electric field, Hall term ($J_p \times B/eN$ (green), $J_b \times B/eN$ (red), where Jp corresponds to current obtained from particle and Jb from curlometer), and $-V_i \times B$ electric field (black). (d) The ratio of the magnitude of the $E + V_i \times B$ and the Hall term to the convection $-V_i \times B$ electric field with the same color scheme as in (c).

The intense upward (tailward) field-aligned currents at MMS embedded in a weaker upward field-aligned current around the center of the flow shear region, where the Hall-effects dominates in the normal component of the electric field ($E_Z$), can be also due to the closure current of the Hall-effects of reconnection (Nakamura et al., 2018). In such case intense tailward field-aligned current is expected in the outflow region of reconnection current sheet (e.g., Ma and Le., 2001) consistent with the observation. Considering that the field-aligned current observed at Cluster has similar intensity, the upward current can also be interpreted as the closure current of the Hall effects of the reconnection region. Since the intense field-aligned current is observed while Cluster is moving equatorward associated with enhancement of Earthward-duskward fast flow, such reconnection region is expected to be formed Earthward (but still tailward of Cluster) and dawnward of the initial source region of the earlier BBF related to the DF.

### 3.3. Electron acceleration

Using data from simultaneous observations of a dipolarization front by MMS and Cluster we examine the electron acceleration properties. We compare the energy distribution around the dipolarization front by 8-s average distribution before (Tb) and after (Ta) the acceleration observed at MMS1 and Cluster4. The time intervals of Tb and Ta are marked in Figures 5e and 5k. They are Tb=14:00:57~14:01:05 UT, Ta=14:01:08 ~14:01:16 UT for MMS and Tb=14:01:09 ~14:01:17 UT, Ta=14:01:17 ~14:01:25 UT for Cluster. Figures 7a and 7b are energy flux distributions observed by MMS and Cluster for perpendicular and field-aligned (parallel or antiparallel) components. To avoid the noise at the higher-energy bins with small counts in burst mode data for MMS, we used data only when at least 50% of data points were available for the averaging. Still it covers at least up to 5 keV, which is more than 10 times the thermal energy (~few 100 eV), thus covering also sufficiently the non-thermal population. As can be seen from Figure 5, the energy distribution after the dipolarization front shifts toward higher energies. The enhancement is more significant for Cluster. We examine whether this difference can be explained by the fact that Cluster was located at the dawnward part of the dipolarization front, whereas MMS was at the duskside edge of the dipolarization front. That is, we check whether both spacecraft were observing plasma from a similar source transported Earthward due to enhanced flux transport rates, but experiencing different acceleration rates due to limited dawn-dusk extent of DFB. If only adiabatic acceleration is considered allowing the Liouville theorem to hold, the phase-space density should be kept constant from the source (for example from the reconnection region located tailward of the spacecraft) to Cluster and MMS. In such case the gained energy is explained by the change in the magnetic field and one can therefore infer the change in phase space density once magnetic field change is known. This method has been successfully applied in previous studies to show features of adiabatic acceleration in a dipolarization event (Smets et al., 1999) and in an injection event (Apatenkov et al., 2007) by using magnetic field model. Here we used this method to compare the phase space density with Cluster and MMS to characterize the acceleration.

Comparison of electron acceleration at MMS and Cluster are made in Figures 7g for the perpendicular distribution. Here we used phase space density profile from the interval Ta for MMS and Cluster (Figure 7f) for comparison. The difference in energy between Cluster and MMS, $\Delta\varepsilon = \varepsilon_{C4} - \varepsilon_{MMS1}$, for different phase space densities value (an example of $\Delta\varepsilon$ is indicated in Figure 7f) is plotted over the energy observed by Cluster, $\varepsilon_{C4}$. The solid curve shows the profile using the observed phase space density curves above 1 keV shown in (Figure 7f). Since the density difference between Cluster and MMS was small (16%), the profile will not change even if the density difference is taken into account, i.e. phase density can be 16% higher at MMS compared to Cluster due to high density, is taken into account as shown in blue-dotted curve, for which MMS phase space density was multiplied by 0.84. The dashed curves are the log-linear fit to the observation (solid curve), which is $\Delta\varepsilon = 0.46\varepsilon^{1.0}$. If we consider only equatorial motion of the electrons convecting Earthward perpendicular to the magnetic field ($B_Z$) experiencing dawnward drift due to gradient B drift opposite to the dawn-dusk electric field (such as illustrated in Birn et al., 2012), the energy difference between MMS and Cluster for a given phase space density (and therefore the energy) should be proportional to the energy and the magnetic field difference, due to the conservation of the first adiabatic moment, $\Delta\varepsilon = \varepsilon \Delta B/B$, as a result of betatron acceleration. That is, the difference in the energy gain then is only due to the difference between the initial and final point of the magnetic field. Assuming that the initial magnetic field was the same at the source for MMS and Cluster, we can use Cluster and MMS observations, i.e. $B_{Z\,C4} = 13.0$ nT and $B_{Z\,MMS1} = 3.3$ nT, to estimate the expected energy difference for the adiabatic acceleration as $\Delta\varepsilon = \Delta B/B\varepsilon = ((B_{Z\,C4} - B_{Z\,MMS1})/B_{Z\,C4})\varepsilon = 0.74\varepsilon$, which is shown as red line. It is interesting to note that the observed energy dependence of $\Delta\varepsilon$ has an overall profile of $\Delta\varepsilon \propto \varepsilon$ as expected for an adiabatic acceleration from the same source, but in a different magnetic flux transport/DFB region. Electrons reaching Cluster experienced a steeper enhancement of magnetic field, which corresponds to the center of the dipolarized flux bundle region. On the other hand, those electrons reaching MMS are originating from the duskside of the DFB where the magnetic field is smaller.

Figure 7h shows the energy difference for field-aligned components. Those field-aligned electrons are bouncing on Earthward collapsing field-lines of DFB and will also drift dawnward due to the curvature drift opposite to the dawn-dusk electric field and become accelerated due to Fermi acceleration of type B (Birn et al., 2011; Birn et al., 2012, and reference therein], which should again depend on the change in magnetic field and on the energy level (i.e. phase space density). The observed curve (solid curve) showed also an almost linear gradient based on the fitted curve (dashed line), $\Delta\varepsilon = 0.28\varepsilon^{1.1}$. The red curve shows the same as Figure 7g as a reference curve showing log linear dependence to the energy. The discrepancy between the observations and the red curve is larger compared to Figure 7g. Although Bz certainly is an indicator also for the field curvature, it may be an oversimplified estimation.

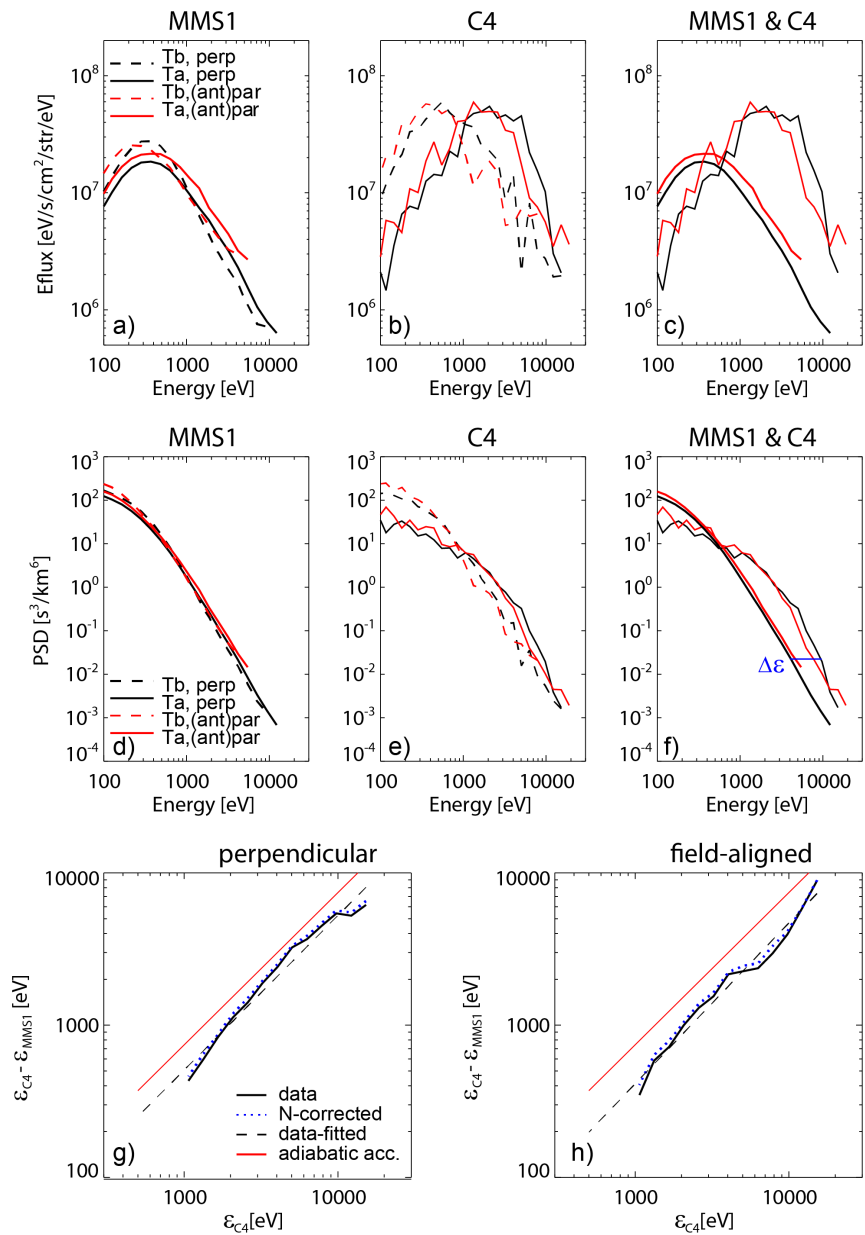

Figure 7. Electron energy distribution before the dipolarization front (Tb) and after the dipolarization front (Ta) observed at MMS1 and Cluster4. Time intervals of Tb and Ta are marked in Figures 5e and 5 k. Energy flux

distribution observed by (a) MMS and (b) Cluster at Tb (dashed curves) and Ta (solid curves), and by (c) MMS and Cluster at Ta. Phase space density obtained by (d) MMS and (e) Cluster at Tb and at Ta, and (f) MMS and Cluster at Ta. Profiles of the perpendicular (field-aligned, i.e. parallel and anti-parallel) component are drawn in black (red) in (a)-(f). Comparison of electron acceleration profile at MMS and Cluster for (g) perpendicular and (h) field aligned electrons by examining the observed and expected difference in energy gain between Cluster and MMS. The solid curve shows the difference in energy between Cluster and MMS using the phase-space density curves shown in (f) for different electron energy at Cluster. The dashed curves are the log linear fit to the observation and the blue dotted curves are the observed curve multiplied by density difference. The red line shows the expected profile for simple adiabatic acceleration (detailed explanation is given in the text).

## 4 Summary and discussions

MMS and Cluster, separated by ~4 $R_E$ in dawn-dusk direction, detected fast flows accompanied by a dipolarization front nearly simultaneously preceding a weak substorm (AL ~-200 nT ) in the premidnight magnetotail at $X$ ~-14 $R_E$. The four MMS spacecraft (and the two Cluster spacecraft in science operation mode) were staying in the southern (northern) hemisphere plasma sheet, enabling a unique constellation of multi-point, multi-scale measurements to determine the changes in the current sheet structure and motion relative to the spacecraft. Knowledge of the relative position to the current sheet was essential to deduce the spatial structure of the flow and the fields, to determine the direction of the flow vortex and to enable reconstruction of the BBF and current sheet evolution from a limited number of spacecraft.

The passage of the dipolarization front could be monitored by Cluster from the dawnside near the center of the flow and by MMS from the duskside edge. The orientation of the front, change in the field direction, and signatures of the thickening of the current sheet, which we could also deduce from the multi-scale gradient measurements, was consistent with a previous statistical study of dipolarization fronts (Liu et al., 2013). Interestingly, this thickening of the current sheet is followed by current sheet thinning in the next minute as shown in Figure 3. Such current sheet thinning has been previously detected by THEMIS spacecraft, P1 and P2, at similar down-tail distance, $X$=-12 ~-14$R_E$, when other spacecraft, P3, P4, and P5, closer to Earth, $X$=-8~-10$R_E$, detected a flow reversed from Earthward to tailward at thickening current (Panov et al., 2010a). Their observation indicates that current sheet thinning can be related to a part of flow bouncing process. While the flow reversal from Earthward to tailward following the current sheet thinning was detected only for MMS but not for Cluster, in contrast to the reversal observation of both P1 and P2 in the THEMIS event (Panov et al, 2010a), the disturbances in $B_x$ of MMS in the southern and Cluster in northern hemisphere are very similar to P1 in the northern and P2 in southern hemisphere, respectively. Global MHD simulation of BBF by Merkin et al. (2019), which successfully reconstructed a MMS observation of near-Earth dipolarization front (Panov et al., 2019), also obtained a stretched plasma sheet behind the deeply penetrated BBF in the dipolar region as a part of the spatial and temporal evolution of the flow vortex with bouncing signatures. The observed thinning may therefore be due to a larger-scale (in radial distance) flow bouncing process and MMS/Cluster could be located at the tailward side of the demarcation line of flow boucing region, i.e. the dipole region closer to Earth where pressure gradient force works most effectively.

Alternative interpretation is that the current sheet thinning is associated with a newly formed reconnection at the wake of DF as showin in the PIC simulation of the thin current sheet with Bz hump (Sitnov et al., 2013, 2017, 2021) or the DFs associated with interchange fingers (Pritchett and Coronitti, 2011, 2013; Pritchett, 2013). In such case Cluster equatorward motion during the current sheet thinning suggests that the new reconnection region should be formed closer to Cluster than the previous BBF source region. Such configuration is indeed expected in the reconnection formed in a thin current sheet just behind the DF for cases where initial flow is due to reconnection (Sitnov et al., 2021) as well as interchange fingers (Pritchett, 2013). Formation of enhanced ion-scale field aligned current regions and Hall-electric field concentrated at the flow shear region, and feature of the bifurcated current sheet indeed suggest that ion-scale physics are important for the current sheet thinning process supporting such interpretation. Yet, it should be noted that the entire scale-size of the dipolarization front in the dawn-dusk dimension is quite wide (>4RE) in our observation, much larger than an ion-scale interchange finger (Pritchett and Coronitti, 2011, 2013). The current sheet thinning therefore maintains both the ion-scale current sheet disturbances and the large-scale thinning (MHD) modified by the large scale flow evolution.

Another important feature during the current sheet thinning is the dawnward widening of the flow burst region as illustrated in Figure 4j, which we only could obtain because of the unique configuration of MMS and Cluster separated in dawn-dusk direction. The flow shear as well as the field-aligned current direction confirm this dawnward evolution. That is, if the thinning is indeed due to flow bouncing, it should also contain some dawn-dusk asymmetric evolution. That is, an initially wider flow channel becoming localized one concentrating at the duskside part so that MMS and Cluster first observed a narrow front followed by a wider flow channel due to some complex evolution mechanism. An alternative way to interpret is to consider that the width of the flow reflects the source region of the flow, i.e. the reconnection region itself, and interpret the change as change in the width of the source region. In Hall-MHD simulations dawn-dusk evolution of the reconnection region has been predicted to take place reflecting the direction of the motion of the initial current sheet carrier for a reconnection without guide field (Nakamura, et al., 2012; Shepherd and Cassak, 2012), which is expected to be the case for the near-Earth magnetotail reconnection. That is, for dawnward evolution the current carrier has to be electrons. We can check whether our observed speed is consistent with such mechanism. From the change in Cluster location, which was initially located on the dawn part of DFB and later at dusk part of DFB, one can infer that the center of BBF is likely moved dawnward at least on a scale of MMS-Cluster separation, i.e. 4 $R_E$, within ~1 minute. That is from the observation one can estimate a dawnward expansion speed of the BBF region, $V_{obs}$ ~ 4 $R_E$ /1 min = 430 km/s. While we have no observations of the initial current sheet in the reconnection site, using the observed initial density n = 1/cc and using lobe field $B_0$ =20~25 nT from Figure 2 and assume that the reconnection took place in a thin current sheet, thinner than the ion inertia length ($d_i$ ~230 km) we can estimate the current density for such curren sheet, j = 70 ~86 nA/m$^2$. Assuming that only electon is carrying the current, we obtain that the velocity of such electron will have Ve = 430 ~ 650 km/s, which is comparable to the dawnward extention velocity of BBF, Vobs. A localized reconnection region formed on the duskside to extend dawnward has been also predicted from statical study of occurrence of reconnection region and flow direction by Nagai et al. (2013). These features are also consistent with the predicted suppression of reconnection toward duskside obtained from 3D reconnection simulation with spatially confined X-line extent (Liu, Y.-H et al., 2019). Considering that our observations are preceding disturbances of a substorm, the observation may show how the initial disturbance is developing to an active thin current sheet region. Still, our six-spacecraft observation provides only a local/temporal observation with respect to the entire BBF evolution. Hence from our observations alone it cannot determined whether a wider flow becomes localized so that MMS and Cluster first observed a narrow front followed by a wider flow region, and/or whether the change in the width originate from changes in the source region

Pi2 activity was visible associated with the flow and dipolarization front disturbances during this event (Figure 1). Yet, there was no clear signatures in the midlatitude magnetogram, i.e. evidence of a developed substorm current wedge. This indicates that the field-aligned current system was too localized to be resolved or the field-aligned current closure took place within the magnetosphere. The observed dawnward evolution of the flow and thin current sheet (Figure 4j), however, is consistent with the direction of the evolution of the substorm current wedge detected from 14:13 UT expanding past midnight (Figure S1). It is also consistent with the location of MMS and Cluster, which observed the dipolarization front with localized magnetic disturbance of wedegelet, at the western sector of the SCW developed dawnward in the course of the substorm.

Adiabatic electron acceleration associated with dipolarization in the magnetotail has been observationally shown by comparing the change in the observed flux with that expected from the change in the magnetic field using magnetic field models (Smets et al., 1999; Apatenkov et al., 2007). Observations are also successfully compared with the results from particle tracing in an MHD simulation of magnetotail reconnection (Birn et al., 2014) and in an analytical model of electric and magnetic field perturbations of a transient, localized DFB (Gabrielse et al., 2016). These models well reconstruct the spacecraft observations of spectral change during DFB passage. By comparing the electron spectra observations of the dipolarization front in dawn-dusk separated location by Cluster and MMS, we showed that our observations are consistent that such acceleration associated Earthward transport from the same source can explain the overall change in the electron spectra difference. In our observation we can directly compare the spectra without any modeling, since we are comparing observation expected from the same BBF event, i.e. same source. Such comparison, however, will only work if the electron energy can be assumed sufficiently small with only slow magnetic field drift so that the electron can stay within the DFB region. As shown by Gabrielse et al. (2016) larger energy electron motion is much more defined by the local gradients in $B_z$ surrounding the DFB itself. They reach the spacecraft from the dusk side of the DFB region due to gradB drift and cannot experience sufficient acceleration due to Earthward motion. It becomes therefore harder to make the assumption that more energetic electrons come from the same source population, which could explain the gradual flattening of the curve in the higher energy range in Figure 7g..Yet, the lower energy profile being consistent with the adiabatic acceleration for this event and confirms our original conjecture that the observed accelerated electrons in the energy range of 1~10 keV are from the same source region of a BBF. That is, althouth there is a more complex 3D evolution in the DF trailing region as discussed before, it did not deform/overtake the front so that the initial signatures of the observable DF seem to be still maintained as a rather mesoscale (MHD scale) simple DF for this event.

## 5. Conclusion

A unique constellation of two multi-point missions, MMS and Cluster, enabled multi-scale observations of localized fast flow and dipolarization front. Based on gradient analysis for both small-scale (less than ion scale) and for meso-scale (a few $R_E$), we could determine the evolution of localized fast flow and dipolarization front by also monitoring the motion/position of the spacecraft inside the dynamically changing current sheet.

MMS and Cluster observed disturbances in flow and fields consistent with a crossing of a dipolarization front/dipolarizing flux bundle (DFB). It is the first conjugate observation of a DF covering both its duskside and dawnside part. This unique near simultaneous observation of a DFB, enabled the identification of electrons accelerated adiabatically in a localized rapid flux transport region in a quantitative way using the spectra differences.

Following the passage of the dipolarization front and subsequent thickening of the current sheet due to the DFB, the current sheet thinned along with further enhancement of flows and the dawnward extension of the BBF region. Intense field-aligned currents were observed at the off-equatorial boundaries of the thin current sheet in the flow shear region, which is likely produced due to small-scale processes at the boundaries.

The observed change in the BBF and current sheet structure may indicate a complex evolution of the Earthward flow channel with flow bouncing signatures as well as the evolution of the source region, such as a dawnward expansion or new activation of the reconnection. Eastward expansion of the SCW obtained from the ground-based observations supports the latter view. Yet, to confirm such temporal and spatial evolution of the complex BBFs, a dawn-dusk as well as radial separation of the spacecraft would be essential, thus we await the realization of a constellation mission including larger scales combined also with small scales.

## Acknowledgments

We thank M. Kubyshkina providing results from magnetic field model. This work was supported by the Austrian Science Fund (FWF): I2016-N20, I3506-N27, P32175-N27 and by the Austrian Research Promotion Agency (FFG): 873685 and by RSF grant 18-47-05001. I. Dandouras thanks CNES for its support to the Cluster project at IRAP. The MMS data are publically available from Science Data Center at CU/LASP (https://lasp.colorado.edu/mms/sdc/public/) and the Cluster data from Cluster Science archive (https://csa.esac.esa.int). Ground based pseudo AE data are available from THEMIS data directory (http://themis.ssl.berkeley.edu/data/themis/thg/).The midlatitude magnetogram data used in this study are available from https://intermagnet.github.io/. We thank the national institutes that support the magnetic observatories collecting those data and INTERMAGNET for promoting high standards of magnetic observatory practice (www.intermagnet.org).

# Thin Curent Sheet Behind the Dipolarization Front

R. Nakamura[1], W. Baumjohann[1], T. K. M. Nakamura[2,1], E. V. Panov[2,1], D. Schmid[1], A. Varsani[1], S. Apatenkov[3], V. A. Sergeev[3], J. Birn[4], T. Nagai[5], C. Gabrielse[6], M. André[7], J. L. Burch[8], C. Carr[9], I. S Dandouras[10], C. P. Escoubet[11], A, N. Fazakerley[12], B. L. Giles[13], O. Le Contel[14], C. T. Russell[15], and R. B. Torbert[8,16]

[1]Space Research Institute, Austrian Academy of Sciences, Graz, Austria

[2]Institute of Physics, University of Graz, Graz, Austria

[3]St Petersburg State University, St Petersburg, Russia

[4]Space Science Institute, Boulder, CO, United States

[5]ISAS/JAXA, Sagamihara, Japan

[6]The Aerospace Corporation, Los Angeles, CA, United States

[7]Swedish Institute of Space Physics, Uppsala University, Uppsala, Sweden

[8]Southwest Research Institute, San Antonio, TX, United States

[9]Imperial College London, London, United Kingdom

[10]IRAP, Université de Toulouse / CNRS / UPS / CNES, Toulouse, France

[11]ESTEC/ESA, Noordwijk, Netherlands

[12]Mullard Space Science Lab., Dorking, United Kingdom

[13]NASA Goddard Space Flight Center, Greenbelt, MD, United States

[14]LPP, CNRS, Observatoire de Paris, Paris, France

[15]University of California Los Angeles, Department of Earth Planetary and Space Sciences, Los Angeles, CA, United States

[16]Univ New Hampshire, Durham, NH, United States

## Contents of this file

Figures S1 to S2

## Introduction

The supporting information contains two figures (Figures S1-S2).

Figure S1 shows the results from the substorm current wedge (SCW) model by Sergeev et al. [2011]. Figure S2 shows the observed magnetic field disturbances at (a) 14:1**2** UT and (b) 14:25 UT together with the modeled ones.

The ground-based data used in the analysis to produce Figures S1and S2 are available from INTERMAGNET network. (https://intermagnet.github.io/).

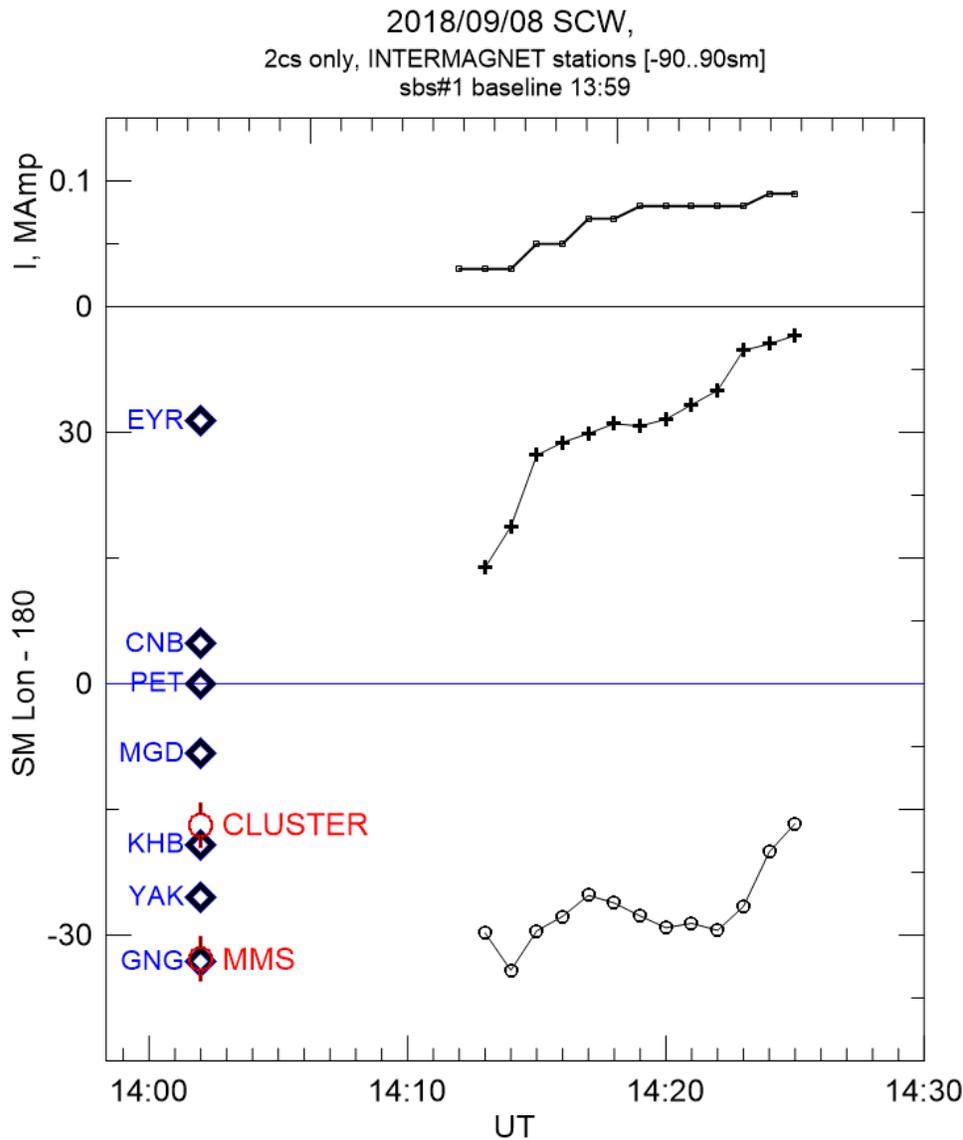

**Figure S1.** Results from the substorm current wedge (SCW) model [Sergeev et al., 2011]. Total current of the SCW (upper panel) and magnetic local time of the upward and downward field-aligned currents (lower panel) obtained by the SCW model using the mid-latitude ground-based magnetic field disturbances. Magnetic local time of Cluster and MMS are also shown. The maximum intensity of the total FAC of SCW was about 0.1 MA during this interval..

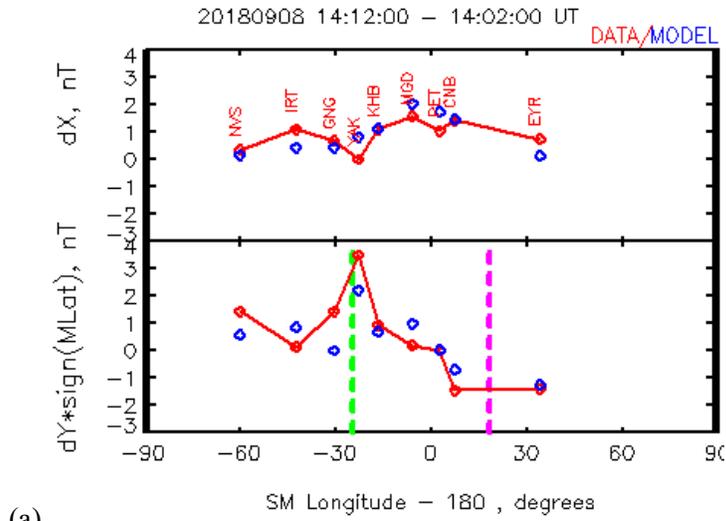

(a)

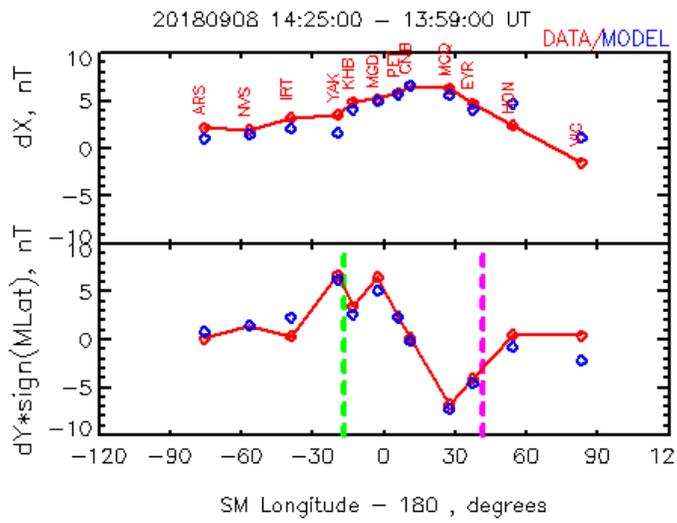

(b)

**Figure S2.** Local time distribution of the disturbances in H(X) and D(Y) components of mid-latitude magnetic fields for (a) 14:12 UT and (b) 14:25 UT. The blue profiles show the modeling result, while the red dots show the data.

3